\documentclass[12pt]{iopart}

\usepackage{iopams}  

\usepackage{graphicx}

\begin{document}

\bibliographystyle{iopart-num}

\title[What do we mean, 'tipping cascade'?]{What do we mean, 'tipping cascade'?}

\author{Ann Kristin Klose$^{1,2}$, Nico Wunderling$^{1,2,3}$, Ricarda Winkelmann$^{1,2}$ and Jonathan F Donges$^{1,3}$}

% adjust affiliations
\address{$^1$ FutureLab Earth Resilience in the Anthropocene, Earth System Analysis \& Complexity Science, Potsdam Institute for Climate Impact Research (PIK), Member of the Leibniz Association, 14473 Potsdam, Germany}
\address{$^2$ Institute of Physics and Astronomy, University of Potsdam, 14476 Potsdam, Germany}
\address{$^3$ Stockholm Resilience Centre, Stockholm University, Stockholm, SE-10691, Sweden}

\eads{\mailto{donges@pik-potsdam.de}, \mailto{akklose@pik-potsdam.de}}
\vspace{10pt}
\begin{indented}
\item[]August 2021
\end{indented}

\begin{abstract}
Based on suggested interactions of potential tipping elements in the Earth's climate and in ecological systems, tipping cascades as possible dynamics are increasingly discussed and studied as their activation would impose a considerable risk for human societies and biosphere integrity. However, there are ambiguities in the description of tipping cascades within the literature so far. Here we illustrate how different patterns of multiple tipping dynamics emerge from a very simple coupling of two previously studied idealized tipping elements. In particular, we distinguish between a two phase cascade, a domino cascade and a joint cascade. While a mitigation of an unfolding two phase cascade may be possible and common early warning indicators are sensitive to upcoming critical transitions to a certain degree, the domino cascade may hardly be stopped once initiated and critical slowing down--based indicators fail to indicate tipping of the following element. These different potentials for intervention and anticipation across the distinct patterns of multiple tipping dynamics should be seen as a call to be more precise in future analyses on cascading dynamics arising from tipping element interactions in the Earth system. 

\end{abstract}

%
% Uncomment for keywords
\vspace{2pc}

% Uncomment if a separate title page is required
%\maketitle
% 
% For two-column output uncomment the next line and choose [10pt] rather than [12pt] in the \documentclass declaration
%\ioptwocol
%

\section{Introduction}
\subsection{The concept of tipping cascades}

Human--induced impacts on the Earth system increasingly endanger the integrity of the Earth’s climate system and some of its most vulnerable components and processes, the so--called tipping elements \cite{lenton2008tipping}. Lately, it has been argued that the risk of potential tipping events or even cascading transitions up to a global cascade is rising under ongoing anthropogenic global warming \cite{steffen2018trajectories,lenton2019climate}. While this is the case, there is considerable debate about the nature of tipping cascades within the scientific community itself and cascading tipping dynamics have been described rather roughly in the recent literature \cite{steffen2018trajectories,lenton2019climate,lenton2020tipping,lenton2013origin,hughes2013multiscale,rocha2015regime,rocha2018cascading,barnosky2012approaching,brook2013does}. 

The term cascade is used in various fields for a certain class of dynamics possibly exhibited by interacting (sub--)systems. It generally describes the sequential occurrence of similar events (event A is followed by event B which is followed by event C etc.). This sequence of events does not necessarily have to be causal opposed to when event A directly causes event B in a domino effect. The notion of a domino effect is sometimes used synonymously to the term cascade. Examples of cascades comprise cascading failures leading to the collapse of power grids as relevant physical infrastructure networks \cite{watts2002simple,buldyrev2010catastrophic,gao2011robustness,gao2012networks,hu2011percolation}. Such a cascade may occur as an initial failure increases the likelihood of subsequent failures \cite{watts2002simple}. In contrast, an initial failure may directly lead to the failure of dependent nodes \cite{buldyrev2010catastrophic}.  

Along these lines, cascading tipping events or regime shifts are increasingly discussed following the rising awareness of a highly interconnected world in the Anthropocene \cite{helbing2013globally}. Tipping elements possibly undergoing a transition into a qualitatively different state after the crossing of some critical threshold were identified e.g. in ecology and climate system science \cite{lenton2008tipping,scheffer2003catastrophic,scheffer2001catastrophic} and comprise, among others, shallow lakes transitioning from a clear to a turbid state \cite{scheffer1989alternative,scheffer1993alternative}, coral reefs \cite{hughes1994catastrophes}, the Atlantic Meridional Overturning Circulation \cite{rahmstorf2005thermohaline,stommel1961thermohaline} and the continental ice sheets on Greenland \cite{robinson2012multistability} and Antarctica \cite{garbe2020hysteresis}. 

In the climate system, multiple interactions between large--scale tipping elements have been identified \cite{kriegler2009imprecise,caesar2018observed,rahmstorf2015exceptional,swingedouw2008antarctic,parsons2019influence,duque2019tipping}. For example, the Atlantic Meridional Overturning Circulation may slow down due to increasing meltwater flux originating from the Greenland Ice Sheet \cite{caesar2018observed,rahmstorf2015exceptional}. Potential drying over the Amazon rainforest basin leading to loss of rainforest resilience may be influenced by the Atlantic Meridional Overturning Circulation \cite{parsons2019influence} on the one hand and the El--Niño Southern Oscillation on the other hand \cite{duque2019tipping}. Rocha et al.~\cite{rocha2018cascading} identified potential links between ecological systems with alternative states such as the interaction of eutrophication and hypoxia or coupled shifts in coral reefs and mangrove systems. 

Tipping interactions do not only exist across different large--scale systems, but span various spatial scales as exemplified by spatially extended (and heterogeneous) ecosystems \cite{lenton2020tipping,rocha2018cascading}. On a local scale, confined ecosystems such as a shallow lake, in fact, consist of discrete units connected through dispersion or other exchange processes with each unit potentially exhibiting alternative stable states \cite{van2005implications,dakos2010spatial,van2015resilience}. Regionally, regime shifts may propagate from one ecosystem entity to the other transmitted, among others, via small streams and rivers \cite{hilt2011abrupt,scheffer2004ecology,van2017regime}, moisture recycling \cite{lenton2020tipping,wunderling2020network,zemp2014importance,zemp2017self} or biotic exchange through e.g. larvae \cite{brook2013does,van2015resilience,scheffer2012anticipating,lundberg2003mobile}. 

Motivated by these and further suggested tipping element interactions, cascading effects arising as potential dynamics have been discussed \cite{steffen2018trajectories,lenton2019climate,lenton2020tipping,lenton2013origin,hughes2013multiscale,rocha2015regime,rocha2018cascading} as a possible mechanism for creating a potential planetary--scale tipping point (of the biosphere) \cite{lenton2013origin,hughes2013multiscale,barnosky2012approaching,brook2013does}. Lenton et al.~\cite{lenton2019climate} stated that we may approach a global cascade of tipping points via the progressive activation of tipping point clusters \cite{schellnhuber2016right} through the increase of global mean temperature. This could potentially lead to undesirable hothouse climate trajectories \cite{steffen2018trajectories}. However, it remains unclear whether and how cascade--like dynamics within the Earth system is promoted by the direction and strength of the existing feedbacks \cite{lenton2020tipping,lenton2013origin,kriegler2009imprecise,wunderling2021modelling}. 

Recently, first conceptual steps \cite{brummitt2015coupled,abraham1991computational} have been undertaken to determine whether the network of Earth system tipping elements is capable to produce global tipping cascades \cite{wunderling2020interacting,gaucherel2017potential}. Using still conceptual, but process--based models, Dekker et al.~\cite{dekker2018cascading} demonstrated a possible sequence of tipping events in a coupled system of the Atlantic Meridional Overturning Circulation and El--Niño Southern Oscillation. Social costs of future climate damages caused by carbon emissions originating from domino effects of interacting tipping elements were studied using an integrated assessment model \cite{lemoine2016economics,cai2016risk}. Earlier, the propagation of critical transitions in lake chains as an ecological example was analyzed, coupling established models of shallow lakes by a unidirectional stream or via diffusion processes \cite{van2005implications,hilt2011abrupt}. The effect of spatial heterogeneity and connectivity of bistable patches on the overall ecosystem response was further studied by the application of simple models for eutrophication and grazing of a (logictically--growing) resource \cite{van2005implications,dakos2010spatial}. In addition, examples beyond the  biogeophysical Earth system possibly giving rise to the propagation of critical transitions were proposed such as coupled subsystems in the fields of economics and finance \cite{lenton2020tipping,brummitt2015coupled}.

\subsection{Descriptions of tipping cascades vary across the literature}
However, tipping cascades or, more generally, patterns of multiple tipping dynamics discussed to arise from the interaction of tipping elements are often loosely described suffering a similar fate as the ancestral ‘tipping point’ concept \cite{van2016you}. We encountered important differences across the description of tipping cascades in the recent literature. These differences are in particular related to whether causality is a necessary ingredient for a cascade or not. For example, the pattern where tipping of one system causes the tipping of another system is described as domino dynamics or tipping cascade by Lenton et al.~\cite{lenton2020tipping}. The propagation of regime shifts by an initial critical transition causing a following one is underpinned by generalized tipping element interactions and termed a cascade by Brummitt et al.~\cite{brummitt2015coupled}. By comparison, the term cascading tipping is used for a sequence of abrupt transitions in Dekker et al.~\cite{dekker2018cascading} that may not necessarily be causal. This notion of cascading tipping is exemplary applied to the Atlantic Meridional Overturning Circulation and El--Nino Southern Oscillation as climatic tipping elements \cite{dekker2018cascading}. Furthermore, and not restricted to causal events, an effect of one regime shift on the occurrence of another regime shift is suggested as cascading in Rocha et al.~\cite{rocha2018cascading} and confirmed to connect ecological regime shifts such as fisheries collapse and transitions of kelp, mangrove and seagrass ecosystems. 

Here we systematically identify and characterize patterns of multiple tipping dynamics such as a domino cascade, a two phase cascade and a joint cascade, which arise in a previously studied system of idealized interacting tipping elements \cite{brummitt2015coupled,abraham1991computational} (section~\ref{sec:res}). In particular, these patterns of multiple tipping dynamics differ in the way of how the critical transition propagates from one tipping element to another. The domino cascade, the two phase cascade and the joint cascade are related to the varying descriptions of tipping cascades in the literature and examples of multiple tipping events with comparable characteristics in the Earth system are given. Furthermore, we address the potential for intervention and anticipation by common early warning indicators based on critical slowing down (see Supplementary Material for details). Implications of the distinct patterns of multiple tipping for the resilience of the Earth system, limitations of studying idealized interacting tipping elements and necessary future research are discussed (section~\ref{ref:disc}).

\section{Patterns of multiple tipping in a model of idealized interacting tipping elements}
\label{sec:res}
In the following, we present distinct patterns of multiple tipping dynamics, which emerge from the linear bidirectional coupling of two idealized tipping elements (figure~\ref{fig:fig_1}, \cite{brummitt2015coupled,abraham1991computational}). Each tipping element depends on its control parameter (or driver), the variation of which may induce a critical transition from a normal to an alternative state with the crossing of a critical control parameter threshold. We consider homogeneous tipping elements, i.e. both tipping elements undergo a critical transition at the same control parameter threshold and on the same intrinsic tipping time scales. A linear coupling term captures the interaction of the tipping elements following Wunderling et al.~\cite{wunderling2020interacting}, where the state of one tipping element is added to the control parameter of another, coupled tipping element. We refer to Wunderling et al.~\cite{wunderling2020interacting} and Klose et al.~\cite{klose2020emergence} for a detailed description of the model of idealized interacting tipping elements. 

The patterns of multiple tipping dynamics described below and illustrated in figure~\ref{fig:fig_2} originate from different pathways through the control parameter space of both tipping elements: The control parameter~$c_2$ of subsystem~$X_2$ as \textit{following} tipping element is kept constant at distinct levels (figure~\ref{fig:fig_2}, going from top to bottom). The control parameter~$c_1$ of subsystem~$X_1$ as \textit{evolving} tipping element is increased (figure~\ref{fig:fig_2}, going from left to right) sufficiently slowly such that this subsystem can follow its (moving) equilibrium. In other words, by a separation of the intrinsic system time scale and the time scale of the forcing, the system can be regarded as a fast--slow system \cite{kuehn2011mathematical}, where the change in the forcing of the system is slow compared to the intrinsic system time scale. We observe the following three qualitatively different dynamic patterns of multiple tipping: 

\begin{figure}[htbp]
	\centering
	\includegraphics[scale = 0.4]{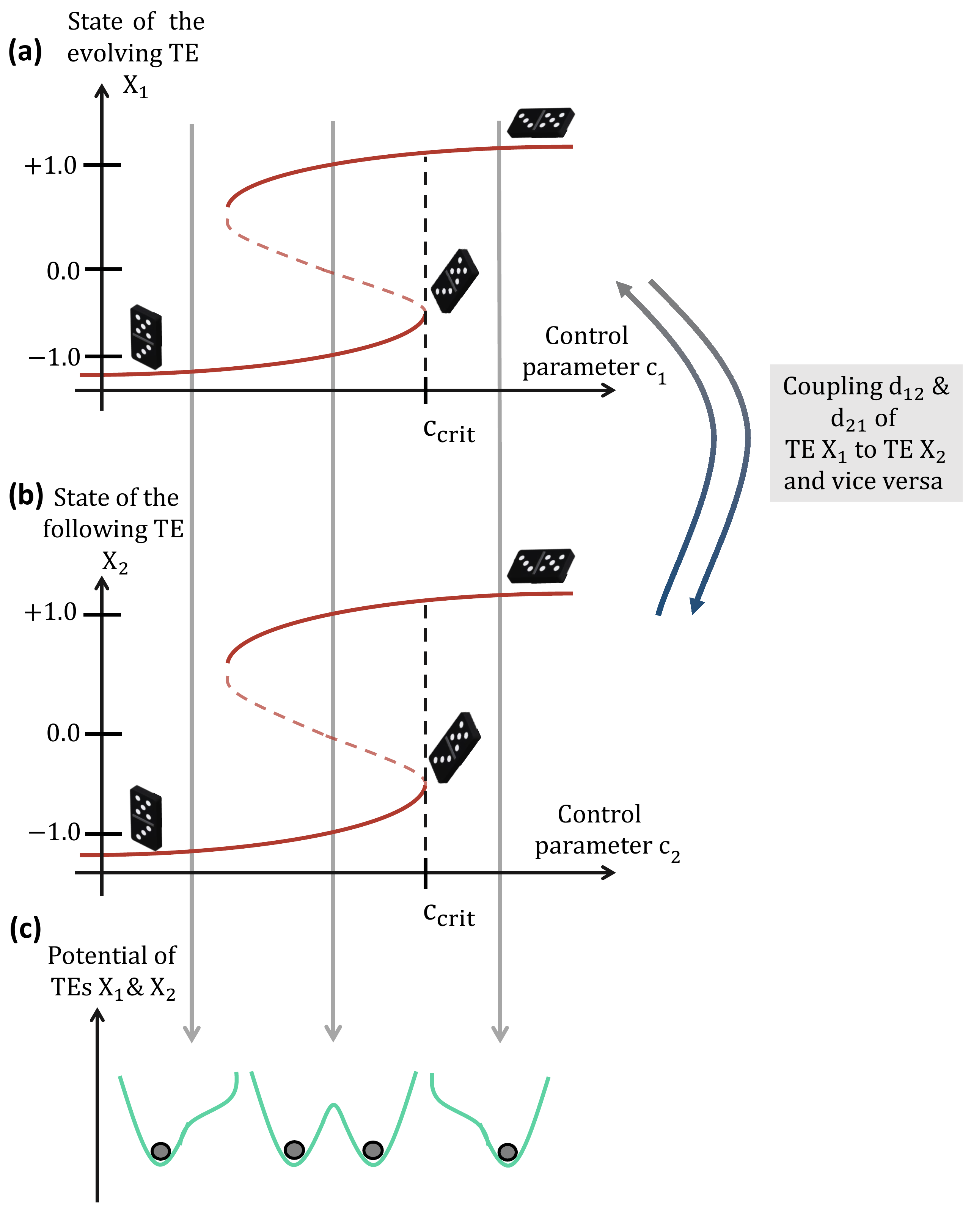}
	\caption{(a) \& (b): Bifurcation diagram of the idealized tipping elements~(TE)~$X_1$ (a) and $X_2$ (b). The respective differential equation is of the form $\frac{\rmd x_1}{\rmd t} =-x_1^3+x_1+c_1+\frac{1}{2}d_{21}(x_2+1)$ for subsystem~$X_1$ and $\frac{\rmd x_2}{\rmd t}=-x_2^3+x_2+c_2+\frac{1}{2}d_{12}(x_1+1)$ for subsystem~$X_2$. Note that for determining the bifurcation diagram of the idealized tipping elements~$X_1$ and $X_2$ the coupling term is not taken into account, i.e. the uncoupled case with $d_{21}= 0$ and $d_{12}= 0$ is shown here. Below the critical threshold~$c_{i_{\rm{crit}}}$, $i = 1,2$, there exist two stable fixed points. As soon as the control parameter~$c_i$ transgresses its critical value~$c_{i_{\rm{crit}}}$, a fold--bifurcation occurs and the system tips from the lower (normal) state~$x_{i^-}^*$ to the upper (alternative) state~$x_{i^+}^*$. (c) Sketch of the potential landscape of the two subsystems in case they do not interact shown as a ball--and--cup diagram.}
	\label{fig:fig_1} 
\end{figure}

\subsection{Two phase cascade (figure~\ref{fig:fig_2}(a))}

An increase of the control parameter~$c_1$ across its threshold and the resulting critical transition of subsystem~$X_1$ is not sufficient to directly trigger a critical transition in subsystem~$X_2$. The system converges intermediately to a stable fixed point (as seen in the phase space portraits) and only a further increase of the control parameter~$c_1$ can initiate the critical transition in subsystem~$X_2$ by the loss of the intermediately occupied stable fixed point. Thus, by limiting the further increase in the control parameter~$c_1$ after the first tipping event of subsystem~$X_1$, a full two phase cascade can be mitigated. We can identify the two phase cascade with the cascade described and simulated in Dekker et al.~\cite{dekker2018cascading}. Within the climate system, a stepwise change in the oxygen isotopic ratio at the Eocene--Oligocene transition may be interpreted as a two phase cascade of the Atlantic Meridional Overturning Circulation as the evolving tipping element and the Antarctic Ice Sheet as the following tipping element in response to a slowly decreasing atmospheric carbon dioxide concentration \cite{dekker2018cascading,tigchelaar2011new}. 

An increasingly slower recovery from perturbations and thus an increase in common statistical indicators such as autocorrelation and variance are observed for subsystem~$X_1$ on the approach of the two phase cascade in a \textit{pre--tipping time span} before the critical transition of subsystem~$X_1$ (Supplementary Material, figure~S1--S3). In contrast, for subsystem~$X_2$, an increasingly slower recovery from perturbations as well as increasing autocorrelation and variance can not be detected in the pre--tipping time span prior to the critical transition of subsystem~$X_1$ (Supplementary Material, figure~S1--S3). However, given the intermediate convergence to a stable fixed point after the critical transition of subsystem~$X_1$ and prior to the critical transition of subsystem~$X_2$, an \textit{intermediate time span} offers the possibility to indicate the upcoming critical transition of subsystem~$X_2$ in the two phase cascade. A step--like change to a relatively higher level of the statistical indicators for subsystem~$X_2$ compared to the respective level in the pre--tipping time span is observed (Supplementary Material, figure~S2--S3, compare also \cite{dekker2018cascading}), indicating an increased vulnerability of subsystem~$X_2$ to a critical transition. The height of the step--like change in the statistical indicators varies with the magnitude of the constant control parameter~$c_2$ as a consequence of an increasingly slower recovery from perturbations in the intermediate time span with increasing magnitude of the constant control parameter~$c_2$. This observation corresponds to the rotation of the eigenvectors and the change in the eigenvalue magnitude of the system of interacting tipping elements, which determine the magnitude and direction of the recovery to perturbations and hence critical slowing down prior to a bifurcation--induced critical transition (\cite{boerlijst2013catastrophic,dakos2018identifying}, Supplementary Material). However, no threshold, i.e. a height of the step--like change above which this following tipping occurs, can be observed but it rather is a continuous and relative quantity. In other words, a step--like change of the statistical indicators (though comparably smaller) may also be present after the critical transition of subsystem~$X_1$ even if a critical transition of subsystem~$X_2$ does not follow. Thus, to use this height of the step--like change to clearly indicate an upcoming following transition may be difficult in practice.

\subsection{Domino cascade (figure~\ref{fig:fig_2}(b))}

For a slightly elevated level of the constant control parameter~$c_2$, the increase of the control parameter~$c_1$ across its threshold and the corresponding critical transition of subsystem~$X_1$ towards its alternative state is sufficient to trigger a critical transition of subsystem~$X_2$. Note that, in contrast to the two phase cascade, no further increase of the control parameter~$c_1$ is necessary to observe the domino cascade, but the tipping of one subsystem (the evolving tipping element) directly causes and initiates the tipping of another (the following tipping element). This corresponds to the description of a tipping cascade given in Lenton et al.~\cite{lenton2020tipping} and Brummitt et al.~\cite{brummitt2015coupled} and the general notion of a domino effect including causality \cite{hornby2015dict}. A notable feature is the expected path of the system in the phase space. Even though the intermediately occupied stable fixed point involved in the two phase cascade is absent, it still influences the dynamics (see phase space, figure~\ref{fig:fig_2}(b)) as a ‘ghost’ (e.g. \cite{strogatz1989predicted,sardanyes2006ghosts,sardanyes2009ghosts,duarte2012chaos}). As demonstrated recently in a conceptual model, domino cascades may propagate through tipping elements in the Earth system, such as the large ice sheets on Greenland and West Antarctica and the Atlantic Meridional Overturning Circulation \cite{wunderling2020interacting, wunderling2020basin}. 

A domino cascade may not be preceded clearly by the increase of the common early warning indicators and relying on these indicators may lead to an unexpected following critical transition of the following tipping element. An increasingly slower recovery from perturbations and thus increasing autocorrelation and variance as common statistical indicators are observed for subsystem~$X_1$ on the approach of the domino cascade in the pre--tipping time span (Supplementary Material, figure~S1--S3). The statistical indicators for subsystem~$X_2$ remain constant though on a relatively higher level than for the two phase cascade in the pre--tipping time span (Supplementary Material, figure~S1--S3). However, no clear intermediate time span prior to the critical transition of subsystem~$X_2$  exists allowing for an additional detection of early warning signals as for the two phase cascade. 

\begin{figure}[htbp]
	\centering
	\includegraphics[scale = 0.5]{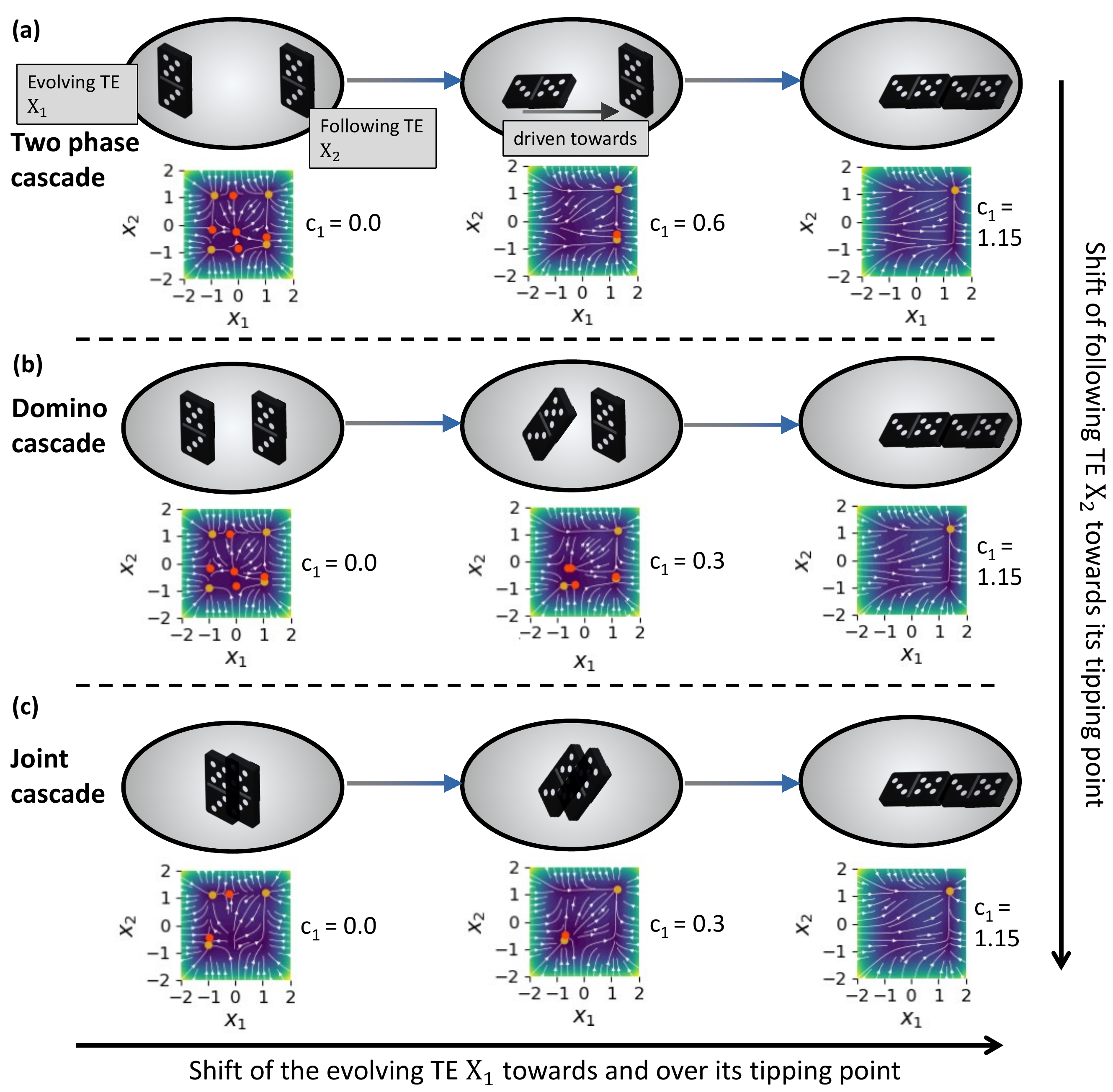}
	\caption{Three different types of tipping cascades depicted as three different situations. From left to right, the critical parameter~$c_1$ of the evolving tipping element~(TE)~$X_1$ is driven closer to and over its tipping point (compare to figure~\ref{fig:fig_1}). From top to bottom, the critical parameter~$c_2$ of the following tipping element~(TE)~$X_2$ is also driven closer to, but not across, its tipping point. In this setting, three different patterns of multiple tipping or cascades can occur. (a) Two phase cascade: The first subsystem~$X_1$ tips and is then shifted closer towards subsystem~$X_2$ by an increase of the control parameter~$c_1$. Then subsystem~$X_2$ tips as well. (b) Domino cascade: The subsystems~$X_1$ and $X_2$ are closer together than in the two phase cascade such that a tipping of subsystem~$X_1$ (middle panel) is sufficient to trigger a critical transition in subsystem~$X_2$. (c) Joint cascade: The two subsystems are very close to each other such that the beginning of a tipping event in subsystem~$X_1$ immediately causes the tipping of the second subsystem~$X_2$ and the tipping events cannot be distinguished. The respective stable fixed point attractors and phase diagrams are shown below the domino sketches. Orange dots represent stable fixed points, while unstable fixed points are given by red dots. The background colour indicates the normalized speed $v = \sqrt{\dot{x}_{1}^2+\dot{x}_{2}^2}/v_{max}$ going from close to zero (purple) to fast (yellow--green).
}
	\label{fig:fig_2} 
\end{figure}

\subsection{Joint cascade (figure~\ref{fig:fig_2}(c))}

Subsystem~$X_1$ and subsystem~$X_2$ may tip jointly with a possible trajectory evolving close to the phase space diagonal for an increase of the control parameter~$c_1$ across its threshold as opposed to the other two multiple tipping patterns. Such a joint cascade is observed with a strongly elevated level of the constant control parameter~$c_2$. The critical transitions of the respective subsystems cannot be clearly distinguished with regard to their order of tipping. Though the case of a joint cascades has not been treated explicitly in the recent literature on interacting tipping elements, a similar behaviour may be observed in spatially extended bistable ecosystems subject to regime shifts \cite{van2005implications,dakos2010spatial}.

For both subsystems, a slower recovery from perturbations is expected prior to their joint tipping (Supplementary Material, figure~S1--S2). For subsystem~$X_1$ autocorrelation and variance increase on the approach of the joint cascade with increasing control parameter~$c_1$. Subsystem~$X_2$ exhibits a relatively high constant level of these statistical indicators prior to the joint cascade corresponding to the level of the constant control parameter~$c_2$ and indicating the vulnerability of this subsystem to critical transitions (Supplementary Material, figure~S3). 

\section{Discussion}
\label{ref:disc}
Studying a system of idealized interacting tipping elements \cite{brummitt2015coupled,abraham1991computational}, qualitatively different dynamic patterns of multiple tipping were identified and characterized as a two phase cascade, a domino cascade and a joint cascade. 

The various patterns of multiple tipping originating from two idealized interacting tipping elements are related to different, though simplified and specific pathways through the control parameter space. In the end, the control parameter evolution determines the emergence of the specific system behavior, which may be a domino cascade, a two phase cascade or a joint cascade. In other words, the control parameter evolution, i.e., the evolution of the forcing, can therefore determine the characteristics of multiple tipping that are observed. However, other factors such as the strength and the sign of coupling are as well decisive for the emergence of tipping cascades. Moreover, in more complex systems, control parameters can not be treated separately for each tipping element and drivers may be shared \cite{rocha2018cascading}.

The different observed patterns of multiple tipping may have implications for the mitigation of tipping by controlling the respective drivers. A limitation of the forcing can prevent the two phase cascade since a critical transition of the evolving tipping element is not sufficient for the spread of a tipping event to a following subsystem. Instead, the critical transition needs to be followed by a further evolution of the respective subsystem’s state before a following critical transition is initiated. However, in a domino cascade an initial critical transition of the evolving tipping element is sufficient to trigger a slightly delayed but inevitable following critical transition of another tipping element.

In addition, the potential success of anticipating the emergence of tipping cascades through early warning indicators based on critical slowing down \cite{wissel1984universal,scheffer2009critical,lenton2011early} was assessed and demonstrated to differ across the patterns of multiple tipping (see Supplementary Material). Using insights of Boerlijst et al.~\cite{boerlijst2013catastrophic} and Dakos~\cite{dakos2018identifying} on critical slowing down in multi--component systems in relation to the eigenvector orientation, it is shown how critical slowing down and common statistical indicators for the anticipation of critical transitions are related to the rotation of eigenvectors and the change in the eigenvalues’ magnitude. Thereby, the analysis of statistical properties of the two phase cascade in Dekker et al.~\cite{dekker2018cascading} is expanded. We find that these common statistical indicators based on critical slowing down may fail for upcoming domino cascades in a system of idealized interacting tipping elements. While increasing autocorrelation and variance are observed for the evolving tipping element on the approach of the domino cascade, constant levels of these statistical indicators were determined for the following tipping element. In the case of a two phase cascade or a joint cascade, the critical slowing down based indicators express some degree of vulnerability (or resilience) in the system of interacting tipping elements. However, their application may be unfeasible in practice. In particular, for the two phase cascade, the critical transition of the evolving tipping element is preceded by increasing autocorrelation and variance of the respective subsystem, while a step--like change towards a relatively higher level of the statistical indicators in the intermediate time span is found for the following tipping element. The joint cascade may be conceivable with a raised but constant level of autocorrelation and variance for the following tipping element accompanied by an increase of statistical indicators for the evolving tipping element. With the slower recovery from perturbations for both tipping elements, correlations between the subsystems’ time series comparable to the application of spatial early warning signals \cite{dakos2010spatial,dakos2011slowing,donangelo2010early,guttal2009spatial,kefi2014early} may unfold. 

As these very specific and simplified scenarios of control parameter evolution demonstrate that an increase of autocorrelation and variance prior to multiple tipping events cannot necessarily be expected, these common early warning indicators should not be relied on as the only way of anticipating cascading critical transitions in systems of interacting tipping elements. Additionally taking into account often referenced limitations, false alarms and false positives in the application of critical slowing down based indicators to individual tipping elements and the anticipation of upcoming critical transitions \cite{boettiger2013early,dakos2015resilience,ditlevsen2010tipping}, it seems to be necessary to invoke a combination of process-–based modelling accompanied by monitoring the system under investigation resulting in predictions as well as data--driven techniques \cite{dakos2015resilience,ditlevsen2010tipping,dakos2012methods} to detect upcoming multiple transitions and, in particular, the domino cascade. 

Note that the presented discussion is restricted to bifurcation--induced tipping with a relatively weak noise and a sufficiently slow change of the tipping element driver is applied. Hence, our examination of tipping cascades excludes early tipping \cite{lohmann2021abrupt} and flickering \cite{dakos2013flickering} due to noise as well as rate--induced effects, which will further influence the presented patterns of multiple tipping, their characteristics such as the intermediate time span of the two phase cascade and hence the potential for anticipation and mitigation. In a related stochastic system, similar patterns were demonstrated as fast and slow domino effects \cite{ashwin2017fast}. The patterns of multiple tipping are expected to change in response to a fast change of the tipping element driver with respect to the intrinsic response time scales, which cannot be ruled out given the current unprecedented anthropogenic forcing of the biogeophysical Earth system \cite{joos2008rates,zeebe2015anthropogenic}. In addition, rate--induced transitions may occur \cite{ashwin2012tipping,wieczorek2011excitability} as suspected based on modelling studies for the Atlantic Meridional Overturning Circulation \cite{alkhayuon2019basin,stocker1997influence,lohmann2021risk}, predator--prey systems \cite{o2019tipping,scheffer2008pulse,siteur2016ecosystems} and for the release of soil carbon in the form of the compost--bomb instability \cite{wieczorek2011excitability,luke2011soil} and may further complicate the early warning of cascading tipping \cite{lohmann2021abrupt,ritchie2016early}. Heterogeneity across the response of tipping elements to the same control parameter level \cite{brook2013does,scheffer2012anticipating} and in the intrinsic time scales of tipping \cite{wunderling2020interacting,ritchie2021overshooting,hughes2013living} was neglected. 

Finally, it is assumed that the long--term behaviour of many real–world systems in terms of the system’s state such as the overturning strength of the Atlantic Meridional Overturning Circulation \cite{stommel1961thermohaline,cessi1994simple}, the ice volume of the Greenland Ice Sheet \cite{levermann2016simple} and the algae density in shallow lakes \cite{scheffer1989alternative,scheffer1993alternative} can be qualitatively captured by the studied idealized tipping elements featuring a fold bifurcation as tipping mechanism. However, biogeophysical and biogeochemical processes involved in the behaviour of these real–-world systems and included in some more complex climate models may either give rise to further types of cascading tipping or may dampen the overall possibilities of tipping behavior \cite{wunderling2020interacting}. 

\section{Conclusion}

Qualitatively different patterns of multiple tipping dynamics in interacting nonlinear subsystems of the climate and ecosystems have been identified in this work. These multiple tipping patterns may emerge as illustrated in a system of idealized interacting tipping elements and include the cases of joint cascades, domino cascades and two phase cascades. As described in Lenton et al.~\cite{lenton2020tipping} and Brummitt et al.~\cite{brummitt2015coupled} as well as corresponding to the general notion of a domino effect \cite{hornby2015dict}, tipping of one subsystem causes or triggers the tipping of another subsystem in a domino cascade. In addition, we find a two phase cascade corresponding to the tipping pattern presented in Dekker et al.~\cite{dekker2018cascading}. While we reveal that it may be possible to find critical slowing down based early warning indicators for the two phase cascade, such indicators can fail in the case of a domino cascade. 

However, our results are limited by the conceptual nature of the system investigated here. In particular, in more complex and process--detailed models of tipping elements the respective nonlinear properties might be smeared out and the presented characteristics of the emerging multiple tipping patterns might be altered due to processes such as strong noise, interactions to other system components or further biogeophysical processes that are not modelled here. % Second, even in case systems can accurately enough be modelled with simple interacting fold-bifurcation based models, the anticipation of multiple tipping cannot always be successful, as discussed above. This also means in a more general way that the common early warning indicators based on critical slowing down should not be relied on as the only measure of anticipating critical transitions. 

Since cascading tipping dynamics have been described rather roughly in the recent literature and the presented patterns of multiple tipping dynamics differ in the potential of their mitigation and anticipation, we suggest to be more precise in future discussions on potential dynamics arising from the interaction of tipping elements and, in particular, on tipping cascades. 
In the future, a quantitative assessment of interacting tipping elements with an ongoing improvement of their representation in complex (climate) models e.g. by including interactive evolving ice sheets into Earth system models \cite{kreuzer2021coupling} as well as the additional use of paleoclimate data \cite{thomas2020tipping} may help to reduce uncertainties on the preconditions for the emergence of tipping cascades and possible early warning indicators based on process--understanding. To the end, these insights may contribute to reflections on the boundaries of the safe--operating space for humanity, and to a better understanding of Earth system resilience with respect to anthropogenic perturbations more generally. 

\section*{References}
\bibliography{ref}

\ack
This work has been performed in the context of the FutureLab on Earth Resilience in the Anthropocene at the Potsdam Institute for Climate Impact Research. J.F.D. and R.W. thank the Leibniz Association for financial support (project DominoES). N.W. acknowledges support from the IRTG 1740/TRP 2015/50122-0 funded by DFG and FAPESP. N.W. is grateful for a scholarship from the Studienstiftung des Deutschen Volkes. J.F.D. an N.W. are grateful for financial support by the European Research Council Advanced Grant project ERA (Earth Resilience in the Anthropocene, grant ERC-2016-ADG-743080). R.W. acknowledges support by the European Union’s Horizon 2020 research and innovation programme under grant agreement no. 820575 (TiPACCs) and no. 869304 (PROTECT). 
\end{document}